\documentstyle[11pt,aaspp4]{article}

\def\etal{{\it et al.}}
\def\eg{{\it e.g.,}}
\def\ie{{\it i.e.,}}

\begin{document}

\title{PG\,1700+518 Revisited:  Adaptive Optics Imaging and a Revised
Starburst Age for the Companion}

\author {Alan Stockton\altaffilmark{1}, Gabriela Canalizo\altaffilmark{1}, 
and Laird M. Close\altaffilmark{2}}
\affil{Institute for Astronomy, University of Hawaii, 2680 Woodlawn
 Drive, Honolulu, HI 96822}

\altaffiltext{1}{Visiting Astronomer, W.M. Keck Observatory, jointly operated
by the California Institute of Technology and the University of California.}

\altaffiltext{2}{Visiting Astronomer, Canada-France-Hawaii Telescope, 
operated by the National Research Council of Canada, the Centre National de
la Recherche Scientifique de France, and the University of Hawaii.}

\begin{abstract}
We present the results of adaptive-optics imaging of the $z=0.2923$ QSO
PG\,1700+518 in the $J$ and $H$ bands.  The extension to the north of the
QSO is clearly seen to be a discrete companion with a well-defined tidal
tail, rather than a feature associated with the host galaxy of PG\,1700+518
itself.  On the other hand, an extension to the southwest of the QSO (seen
best in deeper, but lower-resolution, optical images) does likely
comprise tidal material from the host galaxy.
The SED derived from images in $J$, $H$, and
two non-standard optical bands indicates the presence of dust intermixed
with the stellar component.
We use our previously reported Keck spectrum of the companion, the SED found
from the imaging data, and updated spectral-synthesis models to constrain
the stellar populations in the companion and to redetermine the age of the
starburst.
While our best-fit age of 0.085 Gyr is nearly the same as our earlier 
determination, the fit of the new models is considerably better.
This age is found to be remarkably robust with respect to different
assumptions about the nature of the older stellar component and the effects
of dust.
\end{abstract}

\keywords{galaxies:  interactions, quasars:  individual (PG\,1700+518)}

\section{Introduction}

PG\,1700+518 ($z=0.2923$), one of the more luminous low-redshift
QSOs, shows a bright, arc-like structure extending about 2\arcsec\
to the north of the QSO (Hutchings, Neff, \& Gower \markcite{hut92b}1992; 
Stickel \etal\ \markcite{sti95}1995).  We recently presented a spectrum 
of this extension, showing that it is dominated in the optical by a 
stellar population $\sim10^8$ years old (Canalizo \& Stockton 
\markcite{can97}1997; hereinafter Paper 1).  We argued that
it is plausible that this starburst and the QSO activity were both triggered
by a recent interaction, and that spectroscopic dating of starbursts in QSO
hosts and strongly interacting companions could lead to the development of
an empirical evolutionary sequence for QSOs.  From the imaging data we had 
on PG\,1700+518 at that time, we
were unsure whether the extension was a companion galaxy or a feature
associated with the host galaxy of the QSO.  Here we describe the results of
adaptive-optics (AO) imaging and show that the object is indeed a discrete 
companion galaxy undergoing strong interaction with the QSO host.  We combine
these data with previous imaging in two optical bandpasses
to give a spectral-energy distribution (SED) from 0.42 \micron\ to 1.28 \micron\
in the rest frame in order to place additional constraints on the stellar
populations.  We then use these constraints and
fits of new spectral synthesis models to the Keck LRIS spectrum to obtain a 
more reliable age for the post-starburst component.

\section{Observations and Data Reduction}

The AO observations of PG\,1700+518 were obtained at the f/36 focus of the
Canada-France-Hawaii Telescope (CFHT) on 1997 July 14 and 15 (UT) with
the University of Hawaii AO system (Roddier, Northcott, \& Graves 
\markcite{rod91}1991; Roddier \etal\ \markcite{rod94}1994).  In the version 
of the system used for these observations, an image of the telescope 
primary was formed on a 13-element deformable mirror
that compensated the wavefront aberrations.  Light shortwards of $\sim1$ $\mu$m
was sent by a beamsplitter to a membrane mirror, which was driven at 2.6 kHz
to image extrafocal images on both sides of focus onto an avalanche-photodiode
array.  Corrections for the wavefront errors derived 
from the difference of these extrafocal images were sent to the deformable 
bimorph mirror, which was updated at 1.3 kHz.  
An important feature of this curvature-sensing AO system is that the correction
level degrades gracefully for fainter guide sources.  To approach
the full correction of which the system is capable, a star with 
$R\sim12$ is required.  We used the QSO itself, with $R\sim15.5$, for 
wavefront sensing; nevertheless, we still achieved a substantial degree
of correction.

Light longwards of $\sim1$ $\mu$m was passed by the
beamsplitter to the $1024\times1024$ QUIRC (HgCdTe) infrared camera
(Hodapp \etal\ \markcite{hod96}1996), which gave an image scale of 
0\farcs035 pixel$^{-1}$ and a field of 36\arcsec.  All of the exposures 
were 300 s; we obtained 22 exposures in the $J$-band and 18 exposures in 
the $H$-band.  For both filter sequences, the telescope was offset to a 
new position after every two images in order to allow the construction of
an object-free sky frame.  Flat-field images were obtained from twilight-sky
exposures.  Flux calibration was obtained from observations of the
standard G93-48 (Casali \& Hawarden \markcite{cas92}1992).

The images were reduced by an iterative procedure.  A bad-pixel
mask was made by combining hot pixels from dark frames and dead pixels from
flat-field frames.  Subsets of the dark-subtracted raw images of the target, 
corresponding to $\sim2000$ s integration time each, were normalized by their
median sky values and median averaged to obtain sky frames (bad pixels
being excluded from the calculation).  The sky frames were then scaled to
the sky value of each raw frame and subtracted, and the residuals were
divided by the flat-field frame.  These were then registered to the nearest
pixel and median-averaged.  After a slight smoothing, this rough combined
frame was used to generate object masks for each frame, which were combined
with the bad-pixel mask.  The process was then repeated to generate
better sky frames, better offsets for the alignment, and a new combined
image.  This image was used to replace bad pixels in each flattened frame
with the median of the other frames, so that the bad pixels would not
skew the centering algorithms used to calculate the offsets,  The final
combined image was a straight average of the corrected, sub-pixel-registered,
flat-fielded frames, using a sigma-clipping algorithm in addition to the
bad-pixel mask to eliminate deviant pixel values.
The intrinsic FWHM of the QSO in the summed
$J$-band and $H$-band images are 0\farcs32 and 0\farcs28, respectively,
while the uncorrected image quality averaged $\sim0\farcs8$ at $J$ and
$\sim0\farcs7$ at $H$.

It was not possible to obtain an adequate separate measure of the
point-spread-function (PSF), so we had to generate a model of the PSF
from the QSO itself.  An image of a star obtained the same night with the
same observing setup was found to have a high degree of symmetry, so we
assume elliptical symmetry for the model.  This procedure has the 
disadvantage that it will
prevent the recovery of elliptically-symmetric components of the QSO host
galaxy, but it does allow us to detect and carry out deconvolutions of
non-symmetric components.  We used
the STSDAS tasks {\it ellipse} and {\it bmodel} to create a PSF model
from modified images of the QSO, in which we replaced portions of the
profile near the northern extension with the corresponding portions of
the southern part of the profile, rotated 180\arcdeg\ about the center.
The elliptical fitting routine takes a median average along each elliptical
isophote, reducing sensitivity to azimuthal structure and deviant pixels.

We used this model PSF for both straight PSF subtraction and
deconvolution with {\it plucy} (Hook \etal\ \markcite{hoo94}1994), which 
performs a two-channel deconvolution, using the standard Richardson-Lucy 
algorithm for designated point sources and a Richardson-Lucy procedure 
modified by the inclusion of an entropy term for the remainder of the image.  
The major advantage of {\it plucy} for detecting structure around QSOs is
that the Richardson-Lucy non-negativity criterion is enforced against
the background component instead of against a fixed zero point, and
this feature eliminates the ringing problem typically seen around point
sources in standard Richardson-Lucy restorations.

In order to determine the SED of the companion, we use
optical images from Stockton, Ridgway, \& Kellogg \markcite{sto98}(1998), as
well as the AO $J$ and $H$ images.  We
first convolve the images with appropriate Gaussian profiles to bring
stellar objects to the same FWHM as those in the poorest of the images 
(0\farcs86).  We then subtract similarly convolved PSFs from the QSO
profiles and do the photometry on these subtracted images.

\section{Results and Discussion}

\subsection{The Morphology of the Companion Galaxy}
The AO images of PG\,1700+518 are shown in Fig.\ 1, in original, PSF-subtracted,
and {\it plucy}-restored versions.  
The extension to the north of the QSO,
which has the appearance of an arc-like or ``boomerang'' shape in the best
previous ground-based images (Hutchings \& Neff \markcite{hut92}1992; 
Stickel \etal\ \markcite{sti95}1995; Paper 1), retains that general appearance 
in our higher-resolution images, but it now also shows discrete condensations
within this overall structure.  It is also clearly a separate companion
galaxy:  there is a rather abrupt dropoff in surface brightness just
south of the brightest condensation, whereas we would expect to see more 
continuity with the inner regions if this were a tail associated with the 
host galaxy.  Such a connection should stand out in spite of our using the
QSO to model the PSF, since we expect to be sensitive to any 
non-elliptically-symmetric features.  The companion, though distinct, is 
apparently in the process of merging with the QSO 
host galaxy.  The main features of the companion are consistent between the
$J$ and $H$ images and must be real.  The bright condensation $a$ is
probably the nucleus of the companion.  The apparent tidal tail, curving
to the north and east, contains another condensation, $b$, which may be
a bright star-forming region or even a dwarf galaxy forming from the tidal
debris (\eg\ Duc \& Mirabel \markcite{duc94}1994; Hunsberger, Charlton, 
\& Zaritsky \markcite{hun96}1996).  The nature of features at lower surface
brightnesses is less certain because of the effect of noise on the
deconvolution.  There is clearly material to the east of $a$, looking
like another condensation in the $J$ image, but more like an arc in the
$H$ image.  The peak to the south of $a$ on the $J$ image is apparently
an artifact of inadequacies in the PSF model,
but there is some evidence in these
images for bridge-like material between the companion and the QSO.

If most of the luminous material north of the QSO is associated with
the companion, is there any evidence for tidal debris from the QSO
host?  Stickel \etal\ \markcite{sti95}1995 noted a possible faint extension 
to the SW of the QSO, which can also be seen in Fig.\ 2 of Paper 1.  Here we 
show this feature in two optical bandpasses in Fig.\ 2, where we have 
slightly oversubtracted the wings of the PSF profile to show the SW extension 
more clearly.  We suggest that this is likely a counter-tidal feature from 
the QSO host.  The inner, higher-surface-brightness contours of the host 
galaxy appear to be aligned nearly E--W (Paper 1).

As we were completing this {\it Letter,} we were informed that Hines \etal\
\markcite{hin98}1998 had obtained a Hubble Space Telescope NICMOS image of 
PG\,1700+518.  They also conclude that the extension to the N is a companion,
but they see what we have described as a tidal tail rather as part of a
ring.

\subsection{The SED of the Companion and the Age of the Starburst}

In Paper 1, we presented a spectrum of the companion, corrected for
contamination from the QSO.
We modeled the SED as a superposition of two 
simple stellar populations (instantaneous bursts). 
Our age estimates, based on a $\chi^2$ fit of the Bruzual \& Charlot 
\markcite{bru93}(1993) models to the spectrum of the companion, gave 
0.09 (+0.04, $-0.03$) Gyr and 12.25 Gyr, respectively, for the two components.

Two recent developments encourage us to try to refine these estimates:
better models are now available (Bruzual \& Charlot \markcite{bru98}1998), 
and we now have images in bandpasses covering a wide spectral range,
which can additionally constrain the stellar populations. 
We use the same Keck Low-Resolution Imaging Spectrograph (LRIS) data from Paper 1. 
Because we need fairly high resolution for detailed fitting to features in
the spectrum, we restrict ourselves to the solar-metallicity spectral-synthesis
models based on the Gunn \& Stryker \markcite{gun83}(1983) and Jacoby, Hunter, 
\& Christian \markcite{jac84}(1984) spectra.  We first discuss fits to the 
Keck LRIS
spectrophotometry, considered in isolation; then we show how the models
must be modified to take into account the SED over a wider wavelength region.
Although the fit to the spectrum shown in Fig.\ 3 is from our final model,
it is typical of the quality of fit we find for other models we
discuss here.

If we try a similar approach to that of Paper 1, using only simple stellar 
populations from the newer models (Bruzual \& Charlot \markcite{bru98}1998),
we obtain nearly the same age for the younger population (0.10 Gyr), but
a much younger age for the older population (1.8 Gyr).  While this new
model fits the Keck spectrum much better than did the model given in
Paper 1, and we could no doubt improve the fit even more by adding a 
third, older population, the morphological evidence suggests 
another approach.  The presence of a strong tidal tail
indicates a dominant, pre-existing disk component, and the evidence for recent
star formation indicates a gas-rich system.  Such a galaxy would be expected
to have been forming stars over the entire lifetime of the disk.  We therefore 
consider another range of models:  those in which there has been an 
exponentially-decaying rate of
star formation on which is superposed a single recent starburst.
Specifically, we assume that the younger population can be modelled as a burst
(there is no appreciable difference in this case whether the burst is taken
to be instantaneous or has a finite duration of several Myr)
and that the older population has an exponentially-decaying
star formation beginning 10 Gyr ago.  We find that the age of the younger
population and the quality of the fit are insensitive to the decay rate of 
star formation in the older component over a reasonable
range (time constants $\sim3$--10 Gyr), but that we cannot get as good fits to 
the observed spectrum for either an old instantaneous burst or
a constant star-formation rate.  For definiteness, we assume an exponential 
time constant of 5 Gyr.  Considering only the Keck spectrophotometry, we
find the best $\chi^2$ fit by combining this underlying population with a
starburst with an age of 114 Myr.  We now deal with additional constraints
from the SED at longer wavelengths.

We determine the SED of the companion over a rest-frame range from 0.42 \micron\
to 1.28 \micron\ from the AO $J$ and $H$-band images and images in two
non-standard optical bands centered at 5442 and 7248 \AA, with FWHM of
1002 and 1260 \AA, respectively (see Fig.\ 2; these bandpasses have been
designed to avoid strong emission lines at the redshift of PG\,1700+518).
Figure 4 shows the flux densities in these four filters for a 1\arcsec\
diameter aperture centered 0\farcs4 E and 2\farcs3 N of the QSO.  This
region is far enough from the QSO that the photometry should not be
very sensitive to the chosen scaling or other errors in the PSF 
subtraction process.  While it does not exactly coincide with the
region covered by the Keck LRIS spectrum, and there could be small differences
in the two SEDs, we attempt to find a model that will fit both.
We have explored a wide variety of SEDs involving superpositions of both simple
and composite stellar populations based on the Bruzual \& Charlot
\markcite{bru98}(1998) models; none of these gives a satisfactory fit
both to the Keck LRIS spectrum and to the wide-band SED.  
A pure old population comes moderately close to matching the overall SED,
but it fails completely to reproduce either the strong Balmer absorption
spectrum or the continuum shape of the restframe UV---blue spectrum.  On the
other hand, the ``reddest'' reasonable model we can find that adequately 
fits the 
Keck spectrum and the optical photometry falls well below the $J$ and $H$
points (see Fig.\ 4).
Even this latter model is rather unphysical,
comprising a pure young population (to fit the UV---blue spectrum) and a
pure old stellar population (to attempt to fit the IR photometry), with
nothing in between;
however, the addition of any intermediate-age component
only makes the fit worse.

Clearly one plausible way to raise the relative flux at longer wavelengths
while retaining an early-type spectrum at shorter wavelengths is to
include some sort of reddening due to dust.  However, simply applying
a Galactic reddening law (\ie\ assuming a dust screen between the object
and the observer) is neither appropriate nor very effective.  Screen-like 
reddening sufficient to force a fit to the overall SED results in very
strong variation in extinction across the range of the Keck 
spectral fit, making it difficult to fit simultaneously the lines and continuum.
One needs a means of effecting a substantial reddening of
the IR flux with respect to the optical flux without changing the slope in the
UV---blue region too much.  Witt, Thronson, \& Capuano \markcite{wit92}(1992) 
have shown that dust distributions that
are more-or-less coextensive with the stellar distribution, in addition to
being more realistic, can provide
exactly this sort of reddening.  Such models differ from the standard reddening 
law (due to intervening dust)
in two main respects:  (1) some of the blue light removed along the sightline
is compensated by scattering of blue light emitted in other directions into
the line of sight, and (2) optical depth effects ensure that most
of the light appearing at short wavelengths has suffered little extinction,
while, at longer wavelengths, a larger portion of the total stellar 
distribution contributes to the emergent flux.  We have used the family
of ``dusty galaxy'' models calculated by Witt \etal\ \markcite{wit92}(1992) 
to produce sets of modified Bruzual \& Charlot \markcite{bru98}(1998) models.  
The dust and the stars are both assumed to have constant density within a 
sphere, and the only
variable is the optical depth to the center at a specific wavelength.
While these models are highly artificial, they are sufficient to show the
general nature of the reddening due to embedded dust, and they
give us a good fit to both the LRIS spectrophotometry and the
wide-band SED.  We have included in Fig.\ 4 the model we have found to give
the best fit to both the Keck spectral data and the overall SED.
This same model is shown in Fig.\ 3, both as fit to the Keck spectrum and
as individual components:  the starburst, represented by an 85-Myr-old
instantaneous burst, and the underlying population having star formation
beginning 10 Gyr ago and exponentially decaying with a time constant of 5 Gyr.
Both components are reddened by the curve given by the Witt \etal\ 
\markcite{wit92}(1992) 
``dusty galaxy'' model with a $V$-band central optical depth of 6.

The fit of the model to the Keck LRIS spectrophotometry (Fig.\ 3) is 
remarkably good, reproducing most individual features as well as the
continuum slopes quite accurately.  We can account for most of the
deviations.  Poor fits to H$\beta$ (and, to a lesser extent, H$\gamma$)
absorption are due to distortion of these profiles by general emission
around the QSO not specifically associated with the companion:  there
is both a positive component and a negative component (at a higher
velocity, from the region on the opposite side of the QSO that was used
to subtract off scattered QSO light; see Paper 1 for details).  Similarly, 
the excess peak near
3868 \AA\ is due to [\ion{Ne}{3}] emission.  The broad dip between
4500 and 4600 \AA\ is due to excess subtraction of \ion{Fe}{2} emission,
which is apparently spatially variable (Paper 1).  We have found that we
can improve the fit of the \ion{Ca}{2} $K$ line by including a secondary
burst $\sim2$ Gyr ago, which may be evidence for a previous close passage
of the companion, but this evidence alone is slender enough that we have
chosen not to include this added complexity to the model.  The only
significant discrepancy that we cannot explain is the poor fit to
H10 $\lambda3798$; this may simply be due to a glitch in our observed spectrum. 

\subsection{Towards an Evolutionary Sequence for QSOs}

Although the SED model we have presented is the best fit we have
found to the data,
subject to the constraints of simplicity and astrophysical reasonableness
that we have chosen, we can find other models that fit nearly as well.  However,
in exploring various models, we have found that, for reasonably plausible
combinations of young and old stellar populations and reddening curves 
(\ie\ those that come close to fitting both the LRIS spectrophotometry
and the overall SED), there is very little spread in the age of the young 
population.  We consistently obtain ages in the range from 75 to 100 Myr,
and we believe that this is a reasonable estimate of the uncertainty in
our determination, subject only to any remaining uncertainty in the Bruzual
\& Charlot \markcite{bru98}(1998) models themselves.  We regard this 
robustness in the
age determination as a hopeful sign for our ongoing efforts to develop
an age sequence for triggering events for QSOs lying in the transition
region between ultraluminous IR galaxies and the classical QSO population
in the far-IR two-color diagram (Paper 1; Canalizo \& Stockton 
\markcite{can98}1998; Stockton \markcite{sto98b}1998).

This enthusiasm is only slightly dampened by the fact that the AO imaging
shows clearly that our age determination is for the interacting companion to
PG\,1700+518, rather than for the host galaxy itself.  Obtaining similar
quality spectrophotometry for the extension of the host galaxy to the
SW, which is both considerably fainter and closer to the QSO, would be
extremely difficult.  Nevertheless, its colors appear to be quite similar
to those of the companion, and models of starbursts in interacting pairs
(\eg\ Mihos \& Herquist \markcite{mih96}1996) indicate that star formation 
peaks at times
of closest passage in both participants.  The projected $\sim2$\arcsec\
($\approx7$ kpc) separation between the QSO and companion and the projected
$\sim5$ kpc tail length of the companion are entirely consistent with
a close passage 85 Myr ago, if projection factors are $\sim0.5$ and
mutual velocities are $\sim150$ km s$^{-1}$.  We therefore believe that
the age of starburst in a close, clearly interacting companion is a good
surrogate for an age determined from the QSO host galaxy itself for purposes
of attempting to define an evolutionary sequence.

\acknowledgments

The AO imaging observations would not have been possible without the support of
Fran\c{c}ois Roddier, Malcolm Northcott, and J. Elon Graves.  The AO
system used in these observations was built with the support of NSF
grant AST93-19004.  We thank Jeff Goldader for helpful comments on
the spectral synthesis models.
This research was partially supported by NSF under grant AST95-29078.

\newpage

\begin{figure}
\epsscale{0.7}
\caption{Images of PG\,1700+518 obtained with the Canada-France-Hawaii
Telescope and the University of Hawaii Adaptive-Optics System.  The
left-hand column shows different versions of the $J$-band image, and the
right-hand column shows corresponding versions of the $H$-band image.
The top row shows the unaltered images of the QSO and nebulosity.  The
middle row shows PSF-subtracted versions, where the PSF model has been
derived from the QSO itself.  Any ellipically-symmetric structure in the
QSO host galaxy will be subtracted.  The bottom row shows {\it plucy}
deconvolutions (Hook \etal 1994; see text for details), using the same
PSF model as the deconvolution kernel.  North is up; East to the left.}
\end{figure}

\begin{figure}
\epsscale{0.55}
\caption{Optical images of PG\,1700+518 obtained with the University of
Hawaii 88-inch Telescope.  The filter central wavelengths are given
in the upper-left corner of each panel.  In each case, a PSF has been
slightly oversubtracted in order to show the extensions clearly (the dark
region in the center is the result of saturation in the PSF star, and
the vertical streak below the QSO position is a CCD artifact).  The
insets in the upper left show the images prior to subtraction at the same
contrast as the main panel, and those in the lower right show lower-contrast
versions of the subtracted images.}
\end{figure}

\newpage
\begin{figure}
\plotfiddle{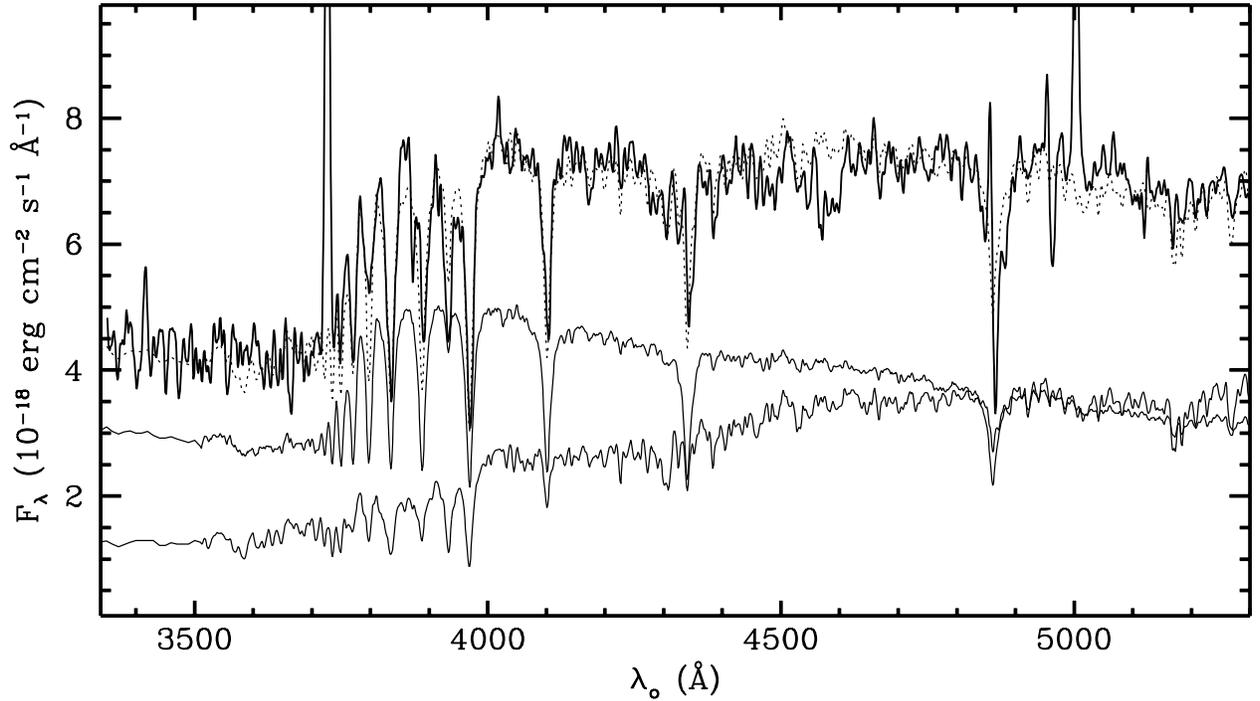}{80mm}{-90}{70}{70}{-270}{400}
\caption{Keck LRIS spectrum of the companion $\sim2\arcsec$ N of PG\,1700+518
(heavy line; see Paper 1 for details).  The dotted line is the best-fit
composite model, comprising an 85 Myr instantaneous burst (upper light line)
and an exponentially decreasing star-formation model with a maximum age of
10 Gyr and a time constant of 5 Gyr (lower light line).  Both components
are multiplied by a reddening curve derived from an embedded-dust model,
as described in the text.  The apparent
[O\,II] $\lambda3727$ and [O\,III] $\lambda5007$ emission lines are not
intrinsic to the companion, but are due to general extended emission around
the QSO.  Weaker, but significant, emission is also present at H$\beta$, 
H$\gamma$, [Ne\,III] $\lambda3868$, and [O\,III] $\lambda4959$.  Similar
emission (but with a higher velocity) in the region used to subtract
scattered light from the QSO distorts the H$\beta$ and H$\gamma$ 
absorption profiles (these negative residuals have been corrected for the
strong [O\,II] and [O\,III] lines).}
\end{figure}

\newpage
\begin{figure}
\epsscale{1.0}
\plotone{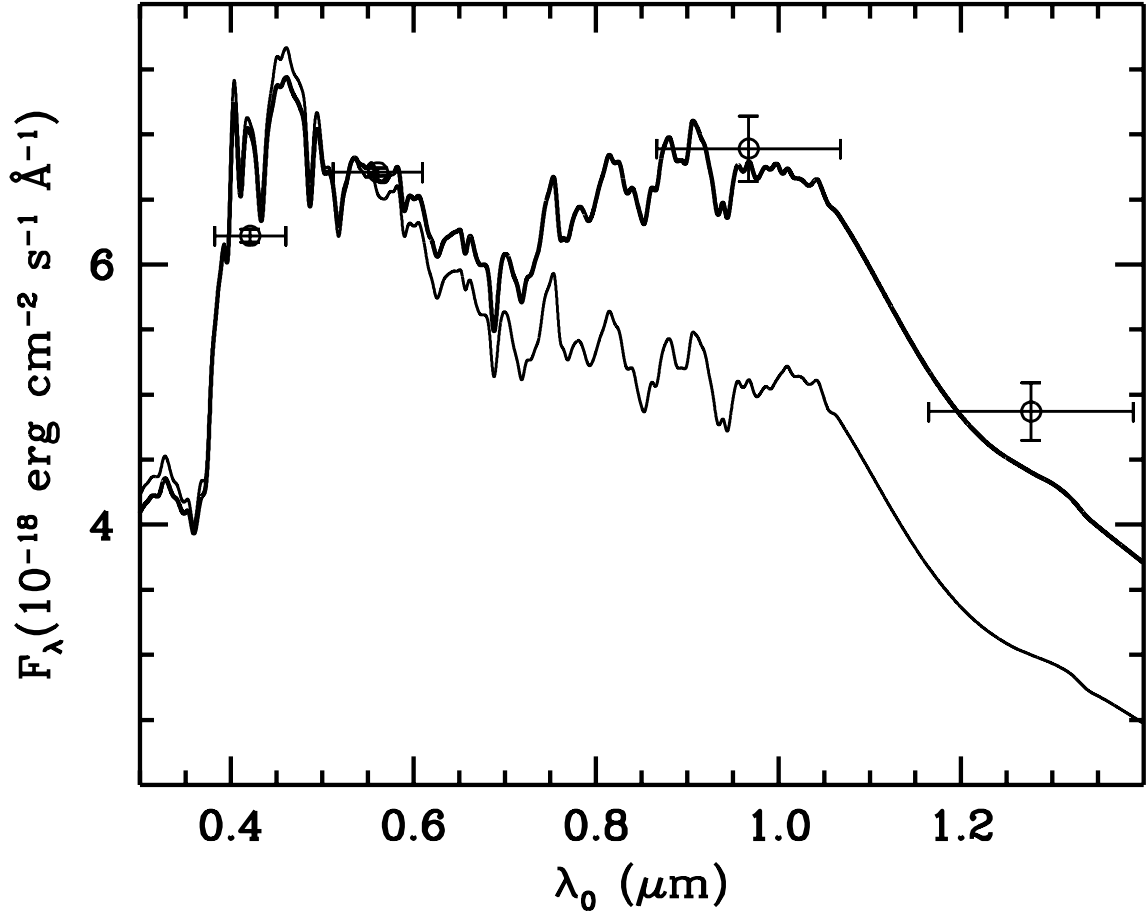}
\caption{Photometry of the PG\,1700+518 companion galaxy.  The lighter
trace shows a model SED comprising instantaneous bursts with ages of
0.114 and 10 Gyrs, and the heavier trace shows the same model as is
shown in Fig.\ 3, renormalized to fit the photometric data.}
\end{figure}

\end{document}